\documentclass[12pt]{article}
\usepackage{amsfonts}
\usepackage{graphics,epsfig}

\def\dalemb#1#2{{\vbox{\hrule height .#2pt
        \hbox{\vrule width.#2pt height#1pt \kern#1pt
                \vrule width.#2pt}
        \hrule height.#2pt}}}

\let\a=\alpha \let\b=\beta \let\g=\gamma \let\d=\delta \let\e=\epsilon
\let\z=\zeta  \let\th=\theta  \let\k=\kappa
\let\l=\lambda \let\m=\mu  \let\x=\xi \let\p=\pi 
\let\s=\sigma \let\t=\tau   \let\c=\chi 
\let\vp=\varphi \let\vep=\varepsilon
\let\w=\omega      \let\G=\Gamma \let\D=\Delta \let\Th=\Theta \let\L=\Lambda
\let\X=\Xi \let\P=\Pi \let\S=\Sigma  \let\Y=\Psi
\let\C=\Chi \let\W=\Omega
\let\la=\label \let\ci=\cite 
  
\def\nn{\nonumber} \def\bd{\begin{document}} \def\ed{\end{document}}
\def\ds{\documentstyle} \let\fr=\frac \let\bl=\bigl \let\br=\bigr
\let\Br=\Bigr \let\Bl=\Bigl
\let\bm=\bibitem
\let\na=\nabla
\def\tU{{\widetilde U}}
\let\pa=\partial \let\ov=\overline
\def\ie{{\it i.e.\ }}
\newcommand{\be}{\begin{equation}}
\newcommand{\ee}{\end{equation}}
\def\ba{\begin{array}}
\def\ea{\end{array}}
\def\ft#1#2{{\textstyle{{\scriptstyle #1}\over {\scriptstyle #2}}}}
\def\fft#1#2{{#1 \over #2}}
\def\F#1#2{{ F_{#1}^{(#2)} }}
\def\cF#1#2{{ {\cal F}_{#1}^{(#2)} }}

\def\={\, =\, }
\def\+{\, +\, }
\def\-{\, -\, }

\def\R{{\bf R}}
\def\sst#1{{\scriptscriptstyle #1}}
\def\oneone{\rlap 1\mkern4mu{\rm l}}
\def\e7{E_{7(+7)}}
\def\td{\tilde}
\def\wtd{\widetilde}
\def\im{{\rm i}}
\newcommand{\ho}[1]{$\, ^{#1}$}
\newcommand{\hoch}[1]{$\, ^{#1}$}
\newcommand{\bea}{\begin{eqnarray}}
\newcommand{\eea}{\end{eqnarray}}
\newcommand{\ra}{\rightarrow}
\newcommand{\lra}{\longrightarrow}
\newcommand{\Lra}{\Leftrightarrow}
\newcommand{\ap}{\alpha^\prime}
\newcommand{\bp}{\tilde \beta^\prime}
\newcommand{\cB}{{\cal B}}
\newcommand{\cO}{{\cal O}}
\newcommand{\vecx}{\vec{x}}
\newcommand{\vecy}{\vec{y}}
\newcommand{\vecp}{\vec{p}}
\newcommand{\vecq}{\vec{q}}
\newcommand{\tr}{{\rm tr} }
\newcommand{\Tr}{{\rm Tr} }
\newcommand{\tchi}{\tilde\chi}
\newcommand{\tpsi}{\tilde\psi}

\newcommand{\cL}{{\cal L}}
\newcommand{\cA}{{\cal A}}
\newcommand{\al}{\alpha}
\newcommand{\cD}{{\cal D}}
\def\sst#1{{\scriptscriptstyle #1}}
\def\0{{\sst{(0)}}}
\def\1{{\sst{(1)}}}
\def\2{{\sst{(2)}}}
\def\3{{\sst{(3)}}}
\def\4{{\sst{(4)}}}
\def\5{{\sst{(5)}}}
\def\6{{\sst{(6)}}}
\def\7{{\sst{(7)}}}
\def\8{{\sst{(8)}}}
\def\ve{\varepsilon}
\def\vf{\varphi}
\def\F{\Phi}
\def\wg{\wedge}

\def \foot {\footnote}
\def \bi{\bibitem}

\def \tr {{\rm tr}}
\def \ha {{1 \over 2}}
\def \td {\tilde}
\def \ci{\cite}
\def \N {{\mathcal N}}
\def \ww {\Omega}
\def \const {{\rm const}}
\def \ss {\sum_{i=1}^3 }
\def \t {\tau}
\def\S{{\mathcal S} }
\def \nn {\nu}
\def \XX {{\rm X}}
\def\CB{{\mathbb C}}
\def\PB{{\mathbb P}}
\def\KB{{\mathbb K}}

\def \lra {\leftrightarrow}
\def \vom {{\bar \omega}}
\def \E {{\mathcal  E}} \def \J {{\mathcal  J}}
\def \YY {{\rm Y}}

\def \d {\del}
\def \rJ {{J}}
\def \sms {sigma models\ }
\def \sm {sigma model\ }
\def \L {\Lambda}
\def \gl {\ell}
\def \tr {{\rm tr\ }}
\def\z{\zeta}
\def\zi{\zeta_1}
\def\zii{\zeta_2}
\def\K{\mbox{K}}
\def\eE{\mbox{E}}   \def \vt {\vartheta}
\def \vr {\varrho}
\def \wup {w}

\def\dg{\dagger}

\def\e{\varepsilon}
\def\p{\phi}
\def\ap{\alpha^\prime}
\def\I{{\cal I}}

\def\R{{\bf R}}
\def\Z{{\bf Z}}
\def\C{{\bf C}}
\def\P{{\bf P}}
\def\xb{{\bar X}}
\def\Tr{{\rm  Tr}}
\def\tr{{\rm  tr}}

\def \del{\partial}
\def\g{\gamma}
\def\s{\sigma}
\def\z{\zeta}
\def\zi{\zeta_1}
\def\zii{\zeta_2}
\def\ov{\over}

\def\I{{\cal I}}
\def\J{{\mathcal J}}
\def \ok {{1\ov \k}}
\def\LL{{\mathcal L }}
\def\RR{{\mathcal R}}
\def\QQ{{\mathcal Q}}
\def \jL {{J}}
\def \om {\omega}
\def \cL {{\mathcal L}} \def \cH {{\mathcal H}}
\def\E{{\mathcal E}}
\def\w{\omega}
\def\b{\beta}
\def\l{\lambda}
\def\eps{\epsilon}
\def\vep{\varepsilon}
\def \De {{\mathcal D}}

 \def \cV {{\cal V}}
\def  \Jt {  {J}_{\rm tot}    }

\def \k {\kappa}
\def\foot{\footnote}
\def \four{{\textstyle {1\ov 4}}}
 \def \third { \textstyle {1\ov 3
}}
\def\det{\hbox{det}}
\def \ci {\cite}

\def \foot {\footnote}
\def \bi{\bibitem}

\def \tr {{\rm tr}}
\def \ha {{1 \over 2}}
\def \tid {\tilde}
\def \vv {{\rm v}}
\def \tl {{\tilde \l}}
\def \XX {{\rm X}}
\def \ta {{\tilde \a}}
\def \fo { {1\ov 4}}
\def \ep {\epsilon}
\def \inti {{\int^{2\pi}_0 {d \sigma \ov 2 \pi}}}

\def \d {\partial}
\def \K {{\rm S}}
\def \el {\ell}
\def \Tr {{\rm Tr}}
\def \P {\Phi}
\def \l  {\lambda}
\def \tl {{\tilde \l}}
\def \bl {{\tilde \l}}
\def \const {{\rm const}}
\def \V {v}

\def \bv {v^*}
\def \vv {{\rm v}}
\def \LL {{\mathcal L}}
\newcommand{\PV}[1]{P_{\!\!_{V_{#1}}}}

\def \bL {\ell}
\def \M {{\mathcal M}}
\def \N {{\mathcal N}}
\def \S {{\rm S}}
\def \vn {\vec n}
\def \tl {\td \l}
\def \td {\tilde}
\def \Prod {\Pi}
\def \O {{\mathcal O}}
\def \Q {{\rm  Q}}
\def \D {\Delta}
\def \N {{\mathcal N}}
\def\tN{{\tilde N}}

\def \m {\mu}
\def \vs {\vec \s}
\def \ie {i.e.}

\def \cD {{\cal D}}

\def \rS {{\rm S}}
\def\as{{\a}}
\newcommand{\bra}[1]{\mbox{$\langle #1 |$}}
\newcommand{\ket}[1]{\mbox{$| #1 \rangle$}}

\newcommand{\auth}{AUTHORS}

\def\thb{\bar{\theta}}
\def\Thb{\bar{\Theta}}
\def\barp{\bar{p}}
\def\barq{\bar{q}}
\def\barc{\bar{c}}
\def\bard{\bar{d}}
\def\e{\epsilon}

\def \bi{\bibitem}
\def \la {\label}

\def \l {\lambda}
\def\foot{\footnote}
\def \tl  {{\tilde \l}}
\def \sql {{\sqrt \l}}
\def \adss {$AdS_5 \times S^5$\ }
\newcommand{\rf}[1]{(\ref{#1})}
\def \ov {\over}

\def\th{\theta}
\def\Th{\Theta}
\def\vth{\vartheta}
\def\btheta{{\bar\theta}}
\def\ttheta{{{\tilde\theta}}}
\def\bttheta{{{\bar\ttheta}}}
\def\vth{\vartheta}
\def\vt{\vartheta}

\def\ra{\rightarrow}
\def\N{{\cal N}}
\def\F{{\cal F}}
\def\uM{\underline{M}}
\def\uN{\underline{N}}
\def\uP{\underline{P}}
\def\cc{\circ}
\def\eqv{\equiv}

\def\ni{\noindent}
\def \ha{{1\ov 2}}
\def \bw {{\rm w}}

\def\r{{\rm r}}

\def \cT {{\cal T}}
\def \no {\nonumber}

\def\a{{\rm\bf a}}
\def\b{{\rm\bf b}}
\def\c{{\rm\bf c}}
\def\d{{\rm\bf d}}

\def\Y{{\rm Y}}
\def\X{{\rm X}}
\def\tY{\tilde{\rm Y}}
\def\tX{\tilde{\rm X}}
\def\dY{\dot{\rm Y}}
\def\dX{\dot{\rm X}}

\def \J {\mathcal{J}}
\def \del {\partial}

\def\dF{\dot{F}}
\def\dG{\dot{G}}
\def\df{\dot{f}}
\def \E {{\cal E}}

\def \S {{\cal S}}
\def \J {{\cal J}}

\def\ms{\mathcal{S}}
\def\mj{\mathcal{J}}
\def\soj{\fr{\ms}{\mj}}
\def \R {{\bf R}}
\def \om {\omega}
\def \tH {\widetilde H}
\def \bE {\bar E}
\def \x {{\cal X}}

 \def \bb {\bar \beta}
\def \W {{\cal E}}

\def \bi{\bibitem}
\def \la {\label}

\def \l {\lambda}
\def\foot{\footnote}
\def \tl  {{\tilde \l}}
\def \sql {{\sqrt \l}}
\def \sqtl {{\sqrt {\tilde \l}}}

\def \HH {{\rm E}}
\def \cS {{\cal S}}
\def \cL {{\cal L}}

\def \adss {$AdS_5 \times S^5$\ }

\def \D {\Delta}
\def \thet {\theta}
 \def \t {\tau}
 \def \p {\del}
 \def \rN {{\rm N}}
 \def\tw{{\tilde w}}
 \def\hJ{{J}}
 \def\hw{{w}}
 \def\hl{{\lambda}}
 \def\hth{{\theta}}
 \def\NN{{\cal N}}
 \def \bv {{ \bar w}}
\def \vn {{\vec n}}
\newcommand{\sfrac}[2]{{\textstyle\frac{#1}{#2}}}
\def \bl {{ \bar \lambda}}

\def \ov {\over}

\def \varpi {{\rm w}}
\def \OO {{\cal O}}

\def \KK {{\rm  K}}
\def \EE {{\rm  E}}

\newcommand{\tT}{\widetilde T}
\newcommand{\tF}{\widetilde F}
\newcommand{\tG}{\widetilde G}
\newcommand{\tJ}{\widetilde J}
\newcommand{\tj}{\tilde j}
\newcommand{\arcsinh}{{\rm arcsinh}}
\newcommand{\arccosh}{\hbox{arccosh}}
\newcommand{\arctanh}{\hbox{arctanh}}
\newcommand{\sech}{\hbox{sech\,}}

\newcommand{\GG}{{\cal G}}
\newcommand{\CC}{{\cal C}}
\newcommand{\BB}{{\cal B}}
\newcommand{\pint}{\makebox[0pt][l]{\hspace{3.4pt}$-$}\int}
\newcommand{\up}{\uparrow}
\newcommand{\down}{\downarrow}
\newcommand{\updown}{\updownarrow}

\def \DD  {{\cal D}}
\def \tDD {\widetilde{\cal D}}
\def \SS  {{ \cal S}}
\def \sn {{\rm sn}}

\def \gt {\frac{g}{\sqrt 2}}
\def \x   {{\rm x}}

\def \te {\theta}

\def \XX {{\rm X}}
\def \rt {{\rm t}}
\begin{document}
\overfullrule=0pt
\parskip=2pt
\parindent=12pt
\headheight=0in \headsep=0in \topmargin=0in \oddsidemargin=0in

\vspace{ -3cm} \thispagestyle{empty} \vspace{-1cm}

\begin{flushright} UUITP-17/06
\end{flushright}

\begin{center}
\vspace{1.01cm}
{\Large\bf
Zero modes for the giant magnon
\vspace{.3cm}
 }

 \vspace{.5cm} {
 J.\,A. Minahan\footnote{joseph.minahan@teorfys.uu.se} \\
 \vskip 0.3cm

\em Institutionen f\"or Teoretisk Fysik, Uppsala Universitet\\
Box 803, SE-751 08 Uppsala, Sweden}

\end{center}

\vskip1cm
 \begin{abstract}
 
 We explicitly construct the eight fermion zero mode solutions for the Hofman-Maldacena giant magnon.   The solutions are naturally gauge fixed under the $\kappa$-symmetry.  Substituting the solutions back into the Lagrangian leads to a simple expression that can be quantized directly.  We also show how to construct the $SU(2|2)\times SU(2|2)$ superalgebra from these zero modes.  For completeness we also find  the four bosonic zero mode solutions.
 
\end{abstract}
\newpage

\renewcommand{\theequation}{1.\arabic{equation}}
 \setcounter{equation}{0}
 \section{Introduction}
 
 There has been much recent progress in our understanding of planar $\NN=4$ gauge theories and its relation to string theory on an $AdS_5\times S^5$ background \cite{Maldacena:1997re,Gubser:1998bc,Witten:1998qj}.  The pivotal  consideration is that the theory has all appearances of being integrable \cite{Minahan:2002ve, Beisert:2003tq,Beisert:2003yb}.  Based on this, Staudacher has advocated that the key to solving the problem is not necessarily to find the full dilatation operator, but instead to concentrate on the $S$-matrix for magnon scattering on very long operators \cite{Staudacher:2004tk}.  
 
 Subsequently, Beisert showed, assuming integrability, that the $S$-matrix is completely determined by the $SU(2|2)\times SU(2|2)$ symmetries, up to an overall phase \cite{Beisert:2005tm,Beisert:2006qh}.  Following this Janik argued that given an involution, which is the analog of crossing symmetry in a relativistic theory, the phase factor would satisfy a deceptively simple looking equation \cite{Janik:2006dc}.  Then in a series of papers, Janik's equation was essentially solved \cite{Beisert:2006zy, Beisert:2006ib, Beisert:2006ez}, where the solutions were shown to be consistent with the known behavior at both strong and weak coupling.  In particular, Beisert, Eden and Staudacher \cite{Beisert:2006ez} showed that a solution consistent with the recent four loop field theory computation in \cite{Bern:2006ew} appeared to be consistent with the phase factor found from string theory \cite{Arutyunov:2004vx, Hernandez:2006tk}.  Further confirmation was given in \cite{Benna:2006nd}, where the authors were even able to predict the two-loop sigma model correction to the $S$-matrix.
 
 Beisert's analysis showed that the dispersion relation for the magnon is consistent with the form
 \begin{equation}
 \eps=\sqrt{1+16g^2\sin^2\frac{p}{2}}\,,
 \end{equation}
 where 
 \begin{equation}
 g=\frac{\sqrt{\l}}{4\pi}\,,
 \end{equation}
 and $p$ is the momentum on the world-sheet for the magnon.   For finite $p$, Hofman and Maldacena explicitly constructed a classical string configuration for which they gave convincing evidence that this was indeed the extension to finite $p$ of the elementary magnon \cite{Hofman:2006xt}.  These so-called giant magnons have a classical energy which is given by $\eps=4g\sin\frac{p}{2}$, where they argued that the world-sheet momentum is related to the stretching of the string on $S^2$, with the maximum value of $p=\pi$ corresponding to a string stretched over the north pole of the sphere and with its ends on the equator.  The results in \cite{Hofman:2006xt} were further generalized to other configurations corresponding to other nonzero $R$-charges \cite{Dorey:2006dq, Chen:2006ge, Arutyunov:2006gs, Minahan:2006bd,Spradlin:2006wk,Bobev:2006fg, Kruczenski:2006pk, Okamura:2006zv,Ryang:2006yq,Kalousios:2006xy}.
 
The Hofman-Maldacena giant magnon is a BPS state, and as such, it should be part of a 16 dimensional short multiplet of $SU(2|2)\times SU(2|2)$.  In terms of the four $SU(2)$ factors, the representation is
$(2,2,2,2)$.    Hofman and Maldacena argued that it would be possible to reproduce this representation if the giant magnon has eight fermion zero modes.

In this paper we will explicitly construct these zero modes from the fermion fluctuation piece of the Green-Schwarz action in the presence of a background giant magnon.   The zero mode solutions turn out to have many simplifications and they can be naturally gauge fixed under the kappa symmetry.  After quantizing these modes, one can then construct the generators of the $SU(2|2)\times SU(2|2)$ subgroup of the full superalgebra.
 
In section 2 we find the fermion zero mode solutions and from these construct the fermionic generators of the $SU(2|2)\times SU(2|2)$ generators.  In section 3 for completeness we also construct the four bosonic zero modes for the giant magnon.  In section 4 we present our conclusions as well as some topics for further study.

\renewcommand{\theequation}{2.\arabic{equation}}
 \setcounter{equation}{0}
 \section{Fermion zero modes}
 
 In this section we explicitly construct the fermion zero mode solutions for the Hofman-Maldacena giant magnon.
 We will work in conformal gauge.  In this case, one finds that the angles on $S^2$ for the giant magnon solution are given by \cite{Hofman:2006xt}
 \begin{eqnarray}
\phi(t,x)&=&t+\vp(x),\qquad\qquad \vp( x)=\arctan(({v\gamma })^{-1}\tanh( x))\nonumber\\
\theta(x)&=&\arccos(\gamma^{-1} \sech( x)) \,,
\end{eqnarray}
where
\begin{equation}
x=\g(\s-vt),\qquad \gamma^2=\frac{1}{1-v^2},
\end{equation}
and $v$ is the velocity of the magnon on the world-sheet.  In terms of the magnon momentum this is
\begin{equation}
v=\cos\frac{p}{2}\qquad\qquad\g^{-1}=\sin\frac{p}{2}\,.
\end{equation}
It is sometimes convenient to write expressions in  terms of $\th$ instead of explicit functions of $x$.  We also introduce a timelike coordinate $\xi$, given by
\begin{equation}
\xi=\g(t-v\s)\,,
\end{equation}
such that $\xi$ and $x$ appear as the boosted variables in the rest frame of the giant magnon.

The action for the fermionic fluctuations about a classical string solution is given by \cite{Metsaev:1998it}
\begin{equation}
S_F=2g\int dt d\s L_F\,,
\end{equation}
where $g\equiv \frac{\sqrt{\l}}{4\pi}$ and the 
Lagrangian $L_F$ has the form
 \bea\label{LF}
 L_F=i\left(\eta^{ab}\delta^{IJ}-\e^{ab}s^{IJ}\right)\;
 \bar{\vartheta}^I\rho_aD_b\,\vartheta^J\;\;,\;\;\ \ \ \rho_a\equiv \G_A
 e_a^A\ , \;\;\;\; e_a^A\equiv E_\m^A(\X)\pa_a \X^\m\ ,
 \eea
where $I,J=1,2,\;s^{IJ}=\mbox{diag}(1,-1),$ \ $ \rho_a$ are
projections of the ten-dimensional Dirac matrices and $\X^\m$ are
 the coordinates of the $AdS_5$  space for $\m=0,1,2,3,4$ and
 the coordinates of $S^5$
for $\m=\phi,\th,7,8,9$. The covariant derivative is given by
 \bea
 && D_a\vartheta^I=\left(\delta^{IJ}{\rm D}_a-\fr{i}2\e^{IJ}\G_*\rho_a\right)\vartheta^J
 \;,\ \ \ \ \quad  \G_*\equiv i\G_{01234}\;,\;\ \G_*^2=1 \ ,
\eea
where ${\rm D}_a=\pa_a+\fr14\w_a^{AB}\G_{AB}, \ \  \w_a^{AB}\equiv
 \pa_a  \X^\mu  \w_\mu^{AB}$.
 In the case of the magnon, one finds that 
 \begin{eqnarray}
 \rho_0&=&\G_0\,+\,\gamma^2(\sin^2\theta-v^2)\csc\theta\,\G_\phi\,-\,v\gamma^2\cot\th\sqrt{\sin^2\th-v^2}\,\,\G_\th
 \nonumber\\
 \rho_1&=&v\gamma^2\cos\th\cot\th\,\G_\phi\,+\,\gamma^2\cot\th\sqrt{\sin^2\th-v^2}\,\,\G_\th
 \nonumber\\
 \om_0&=&-\gamma^2\cot\th\csc\th(\sin^2\th-v^2)\nonumber\\
 \om_1&=&-v\gamma^2\cos\th\cot^2\th\,.
 \end{eqnarray}
 
 The equations of motion derived from \rf{LF} are
 \begin{eqnarray}\label{eom}
 (\rho_0-\rho_1)(D_0+D_1)\vartheta^1&=&0\nonumber\\
 (\rho_0+\rho_1)(D_0-D_1)\vartheta^2&=&0\,.
 \end{eqnarray}
If we change variables to $\xi$ and $x$ in \rf{eom}, then we can reexpress the equations of motion as
 \begin{eqnarray}\label{eom2}
  (\rho_0-\rho_1)\left((1-v)\g(\DD+\p_\xi)\,\vartheta^1-\frac{i}{2}\G_*(\rho_0+\rho_1)\vartheta^2\right)&=&0\nonumber\\
   (\rho_0+\rho_1)\left((1+v)\g(\tDD-\p_\xi)\,\vartheta^2-\frac{i}{2}\G_*(\rho_0-\rho_1)\vartheta^1\right)&=&0\,,
   \end{eqnarray}
   where
\begin{eqnarray}
\DD&=&\p_x+\frac{1}{2}G\,\G_{\phi\th}\nonumber\\
\tDD&=&\p_x+\frac{1}{2}\tG\,\G_{\phi\th}\nonumber\\
G&=&\gamma\cot\th\csc\th (\sin^2\th+v)=\sech\, x\left(1+\frac{v}{\tanh^2 x+v^2\sech^2x}\right)\nonumber\\
\tG&=&-\gamma\cot\th\csc\th (\sin^2\th-v)=-\,\sech\, x\left(1-\frac{v}{\tanh^2 x+v^2\sech^2x}\right)\,.
\end{eqnarray}

It is now straightforward to verify that
\begin{eqnarray}
(\rho_0-\rho_1)^2=(\rho_0+\rho_1)^2=0
\end{eqnarray}
and
\begin{eqnarray}\label{comm}
[(\rho_0-\rho_1),\DD\,]=[(\rho_0+\rho_1),\tDD\,]=0\,.
\end{eqnarray}
We can then define new Dirac fields $\Psi^1\equiv i(\rho_0-\rho_1)\vartheta^1$, $\Psi^2\equiv i(\rho_0+\rho_1)\vartheta^2$.  Moreover, there are another set of nilpotent operators
\begin{equation}
(\bar\rho_0+\rho_1)^2=(\bar\rho_0-\rho_1)^2=0,
\end{equation}
where $\bar\rho_0=\G_*\rho_0\G_*=-\rho_0^\dag$.  If we subtract the first set of nilpotent operators from the second and divide by 2, we are left with the nonsingular operator $\bar\rho_0-\rho_0$ whose square is
 \begin{equation}\label{rhosq}
   (\bar\rho_0-\rho_0)^{2}=4\gamma^2(\sin^2\th-v^2)=4\tanh^2x\,.
   \end{equation}
So $\rho_0\pm\rho_1$ have precisely 8 zero eigenvalues and we can treat the $\Psi^i$ as the fields fixed under $\kappa$-symmetry.  We can then write \rf{eom2} as
 \begin{eqnarray}\label{eom3}
 (1-v)\g(\DD+\p_\xi)\,\Psi^1-\frac{i}{2}\G_*(\bar\rho_0-\rho_0)\Psi^2&=&0\nonumber\\
  (1+v)\g(\tDD-\p_\xi)\,\Psi^2-\frac{i}{2}\G_*(\bar\rho_0-\rho_0)\Psi^1&=&0\,,
   \end{eqnarray}
   where we used the fact that $\rho_1\Psi^1=\rho_0\Psi^1$ and 
   $\rho_1\Psi^2=-\rho_0\Psi^2$.  
   
The zero mode solutions have $\p_\xi\Psi^{1,2}=0$.  Using the relation in \rf{rhosq},
  we can substitute the second equation into the first, and after some further manipulation where we use the relation
\begin{equation}
\tDD\,\frac{\bar\rho_0-\rho_0}{\sqrt{\sin^2\th-v^2}}=\frac{\bar\rho_0-\rho_0}{\sqrt{\sin^2\th-v^2}}\,\DD\,,
\end{equation}
we get  for the zero modes  the compact expression
\begin{equation}\label{compact}
\left(\frac{1}{\tanh x}\,\DD\right)^2\Psi^1-\Psi^1=0\,.
\end{equation}
The solutions of this equation are solutions to either
\begin{eqnarray}
\left(\p_x+\frac{1}{2}G\,\G_{\phi\th}-\tanh x\right)\Psi^1&=&0
\end{eqnarray}
or
\begin{eqnarray}
\label{norm}
\left(\p_x+\frac{1}{2}G\,\G_{\phi\th}+\tanh x\right)\Psi^1&=&0
\end{eqnarray}
but it turns out that only the solutions to \rf{norm} are normalizable.  Because of \rf{comm}, \rf{norm} can also be written as
\begin{equation}
i(\rho_0-\rho_1)\left(\p_x+\frac{1}{2}G\,\G_{\phi\th}+\tanh x\right)\vartheta^1=0\,.
\end{equation}
In other words, we can first find the solutions to \rf{norm} assuming that there is no projection and then fix the $\kappa$-symmetry by projecting  at the very end.  

If we let $\vartheta^1=\vartheta^1_++\vartheta^1_-$ with $\G_{\phi\th}\,\vartheta_{\pm}^1=\pm i\,\vartheta^1_{\pm}$, then the solutions are
\begin{equation}
\vartheta^1_{\pm}(x)=\frac1{4\sqrt{1-v}}\,\sech x\,e^{\pm i\chi}\, U_{\pm}\,,
\end{equation}
where $U_{\pm}$ is constant in $x$ and $\G_{\phi\th}\,U_\pm=\pm i\,U_\pm$.   The phase $e^{i\chi}$ is given by
\begin{equation}
e^{i\chi}=\left(\frac{\sinh x+ iv}{\sinh x- iv}\right)^{1/4}\left(\tanh x+i\,\sech x\right)^{1/2}\,,
\end{equation}
and the prefactor $\frac1{4\sqrt{1-v}}$ has been put in for later convenience.
 Since the spinors are Majoranna-Weyl, we require that $U_-={U_+}^*$. 
The projection then gives
\begin{equation}\label{psi1eq}
\Psi^1=\frac{i}{4\sqrt{1-v}}\,\sech{ x} \left(\left(e^{i\chi}\Gamma_0+e^{-i\chi}\Gamma_\phi\right)U_++\left(e^{-i\chi}\Gamma_0+e^{+i\chi}\Gamma_\phi\right)U_-\right)\,.
\end{equation}
Writing this expression in terms of the Majoranna-Weyl spinor $U=\frac{1}{\sqrt{2}}(U_++U_-)$, we find
\begin{equation}\label{psi1eq2}
\Psi^1=\frac{i}{2\sqrt{2(1-v)}}\,\sech{ x} \Big(\Gamma_0(\cos\chi+\sin\chi\Gamma_{\phi\th})+\Gamma_\phi(\cos\chi-\sin\chi\Gamma_{\phi\th})\Big)U\,.
\end{equation}
Using \rf{psi1eq}, \rf{eom3} and \rf{norm}, we find that $\Psi^2$ is
\begin{equation}
\Psi^2=-\,\frac{1}{4\sqrt{1+v}}\,\sech{ x}\, \Gamma_*\Gamma_\phi\left(\left(e^{i\tchi}\Gamma_{0}+e^{-i\tchi}\Gamma_\phi\right)U_++\left(e^{-i\tchi}\Gamma_{0}+e^{+i\tchi}\Gamma_\phi\right)U_-\right)\,,
\end{equation}
where the phase $e^{i\tchi}$ is given by
\begin{equation}
e^{i\tchi}=\left(\frac{\sinh x- iv}{\sinh x+ iv}\right)^{1/4}\left(\tanh x+i\,\sech x\right)^{1/2}\,.
\end{equation}
In terms of $U$ this becomes
\begin{equation}\label{psi2eq2}
\Psi^2=-\,\frac{1}{2\sqrt{2(1+v)}}\,\sech{ x}\, \Gamma_*\Gamma_\phi\Big(\Gamma_0(\cos\tchi+\sin\tchi\Gamma_{\phi\th})+\Gamma_\phi(\cos\tchi-\sin\tchi\Gamma_{\phi\th})\Big)U\,.
\end{equation}

Let us now substitute these zero mode solutions back into the Lagrangian, where we assume that $U$ can depend on $\xi$.  We then find
\begin{eqnarray}\label{LF0eq}
L_{F,0}&=&-i\,\gamma\,(1-v)\,{\vartheta^1}^\dag\Gamma_0(\rho_0-\rho_1)\p_\xi\vartheta^1-i\, \gamma\,(1+v)\,{\vartheta^2}^\dag\Gamma_0(\rho_0+\rho_1)\p_\xi\vartheta^2\nonumber\\
&=&\frac{i \gamma}{2}\,(1-v)\,{\Psi^1}^\dag\p_\xi\Psi^1+\frac{i \gamma}{2}\,(1+v)\,{\Psi^2}^\dag\p_\xi\Psi^2\,,
\end{eqnarray}
where we used the relations
\begin{eqnarray}
\Gamma_0(\rho_0-\rho_1)&=&-\frac12(\rho_0-\rho_1)^\dag(\rho_0-\rho_1)\nonumber\\
\Gamma_0(\rho_0+\rho_1)&=&-\frac12(\rho_0+\rho_1)^\dag(\rho_0+\rho_1)\,.
\end{eqnarray}
If we now substitute \rf{psi1eq2} and \rf{psi2eq2} into \rf{LF0eq}, we find the remarkable simplification
\begin{equation}
L_{F,0}=-\,i\,\frac{\g}{8}\,\sech^2x\,U^T(\Gamma_0+\Gamma_\phi)^T(\Gamma_0+\Gamma_\phi)\,\p_\xi U\,.
\end{equation}
Integrating $L_{F,0}$ over $x$ then gives the zero mode action
\begin{equation}
S_{F,0}=2g\int d\xi\left(-\,i\,\frac{\g}{4}\,U^T(\Gamma_0+\Gamma_\phi)^T(\Gamma_0+\Gamma_\phi)\,\p_\xi U\right)\,.
\end{equation}

Clearly, only those modes in $U$ which satisfy the light-cone condition $(\Gamma_0-\Gamma_\phi)\,U=0$ contribute.  Since $(\Gamma_0-\Gamma_\phi)$ is null with maximal rank, this means that there are 8 real zero modes.   It is convenient to write $U$ in terms of the $SU(2|2)\times SU(2|2)$ representations preserved by the light cone condition.  To this end, we can define the bispinor components $U_{\al a}$, $\tU_{\dot\al \dot a}$ where the $\al$ and $\dot\al$ spinor indices correspond to the $SO(4)\simeq SU(2)\times SU(2)$ isometry group in the transverse part of $AdS_5$ and the $a$ and $\dot a$ spinor indices correspond to the $SO(4)$ isometry group in the transverse piece of $S^5$.  Quantization of these modes leads to the anti-commutators
\begin{eqnarray}\label{Uanticomm}
\{U_{\al a},U_{\beta b}\}&=&\frac{1}{2g\g}\,\ve_{\al\beta}\ve_{ab}\nonumber\\
\{\tU_{\dot\al\dot a},\tU_{\dot\beta\dot b}\}&=&\frac{1}{2g\g}\,\ve_{\dot\al\dot\beta}\ve_{\dot a\dot b}\nonumber\\
\{U_{\al a},\tU_{\dot\beta\dot b}\}&=&0\,.
\end{eqnarray}
Hence, the bispinor states built from these zero modes are the bosonic states $|Y_{a\dot a}\rangle$ and  $|Z_{\al\dot\al}\rangle$ and the fermionic states $|\Psi_{\al\dot a}\rangle$ and $|\Upsilon_{a\dot\al}\rangle$.

As was argued by Beisert, these states fall into the 16 dimensional representation of $SU(2|2)\times SU(2|2)$, where the states transform under the 4 dimensional short multiplet for each $SU(2|2)$ \cite{Beisert:2005tm,Beisert:2006qh}.  Furthermore, the $SU(2|2)\times SU(2|2)$ superalgebra with central extension can be derived from light cone string theory with the level matching condition relaxed \cite{Arutyunov:2006ak}. 
It is thus of interest to see how the superalgebra arises from the zero modes. 

 The $SU(2|2)$ superalgebra has the six bosonic generators $\RR_{ab}$ and $\LL_{\al\beta}$ and the eight fermionic generators $\QQ_{\al a}$ and $\SS_{a\al}$.  The anticommutators  of the fermionic generators are
\begin{eqnarray}\label{supalg}
\{\QQ_{\al a},\SS_{b\beta}\}&=&\ve_{\al\beta}\RR_{ab}-\ve_{ab}\LL_{\al\beta}-\ve_{\al\beta}\ve_{ab}\,\CB\nonumber\\
\{\QQ_{\al a},\QQ_{\beta b}\}&=&\ve_{\al\beta}\ve_{ab}\,\PB\nonumber\\
\{\SS_{a\al},\SS_{b\beta}\}&=&\ve_{\al\beta}\ve_{ab}\,\KB\,.
\end{eqnarray}
$\CB$, $\PB$ and $\KB$ are central charges.  The smallest nontrivial representation of $SU(2|2)$ is 4 dimensional, with two bosonic and two fermionic states.   Consistency of the algebra requires that the central charges for this representation satisfy the  shortening condition
\begin{equation}\label{cgrel}
\CB^2-\PB\KB=\frac{1}{4}
\end{equation}
The states in the representation are $|\phi_a\rangle$ and $|\psi_\al\rangle$ and the action of the fermionic generators on these states is
\begin{eqnarray}\label{abcd}
\QQ_{\al a}|\phi_b\rangle=\a\, \ve_{ba}|\psi_\al\rangle\qquad&&\qquad \QQ_{\al a}|\psi_\beta\rangle=\b\, \ve_{\al\beta}|\phi_a\rangle\nonumber\\
\SS_{a\al}|\phi_b\rangle=\c\,\ve_{ab}|\psi_\al\rangle\qquad&&\qquad\SS_{a\al}|\psi_\beta\rangle=\d\,\ve_{\beta\al}|\phi_a\rangle\,,
\end{eqnarray}
where $\a\d-\b\c=1$.  The superalgebra in \rf{supalg} is also invariant under the $SL(2)$ outer automorphism \cite{Hofman:2006xt,Beisert:2006qh}
\begin{equation}
\left(\begin{array}{c}\QQ_{\al a}\\ \SS_{a\al}\end{array}\right)\rightarrow\left(\begin{array}{cc}A&B\\ C&D\end{array}\right)\left(\begin{array}{c}\QQ_{\al a}\\ \SS_{a\al}\end{array}\right)\,.
\end{equation}

The central charge $\CB$ is half the energy of the magnon\cite{Beisert:2005tm} and is given by
\begin{equation}
\CB=\frac{1}{2}\sqrt{1+16g^2/\gamma^2}\,.
\end{equation}
The other central charges are
\begin{equation}
\PB=\frac{2g\,\zeta}{\gamma}\qquad \KB=\frac{2g}{\gamma\,\zeta}\,,
\end{equation}
where for a single magnon $\zeta$ is an otherwise undetermined parameter which can be removed by a rescaling\,\footnote{If more than one magnon is present then for the $i^{\rm th}$ magnon \cite{Beisert:2005tm}, $$\zeta_i=\zeta_0\prod_{j<i}e^{ip_j/2}\prod_{j>i}e^{-ip_j/2}\,,$$ where $\zeta_0$ is a common factor for all magnons}.  The $SL(2)$ outer automorphisms will transform these charges while maintaining the relation in \rf{cgrel}.

To write the generators $\QQ_{\al a}$ and $\SS_{a\al}$ in terms of $U_{a\al}$ we need four other generators.  There are  such generators with the correct index structure, namely $(-1)^FU_{a\al}$, where $F$ refers to fermion number for the undotted indices, that is, $(-1)^FU_{a\al}=-U_{a\al}(-1)^F$ while $(-1)^FU_{\dot a\dot \al}=+U_{\dot a\dot\al}(-1)^{ F}$  (The dotted indices have their own fermion number operator, $\dot F$).   The identification of $\QQ_{\al a}$ and $\SS_{a\al}$ with $U_{a\al}$ is basis dependence.   A symmetric identification is
\begin{eqnarray}\label{QSeq}
\QQ_{\al a}&=&\zeta^{1/2}\left(A-B(-1)^F\,\right)U_{a\al}
\nonumber\\
\SS_{a\al}&=&-\zeta^{-1/2}\,\left(A+B(-1)^F\,\right)U_{a\al}\,,
\end{eqnarray}
where 
\begin{eqnarray}\label{ABeq}
A=\frac{2g}{\sqrt{2}}\left(\sqrt{\frac{\gamma^2}{16g^2}+1}\,+\,1\right)^{1/2}\qquad
B=\frac{2g}{\sqrt{2}}\left(\sqrt{\frac{\gamma^2}{16g^2}+1}\,-\,1\right)^{1/2}
\end{eqnarray}
reproduces the algebra in \rf{supalg}.  The generators for the other $SU(2|2)$ can be constructed from $U_{\dot a\dot\al}$ in an analogous fashion.  From the anticommutation relations in \rf{Uanticomm} we have that the action of $U_{a\al}$ on the states is
\begin{equation}
U_{a\al}|\phi_b\rangle=\frac{1}{\sqrt{2g\g}}\ve_{ba}|\psi_\al\rangle\qquad\qquad U_{a\al}|\psi_\beta\rangle=\frac{1}{\sqrt{2g\g}}\ve_{\al\beta}|\phi_b\rangle\,.
\end{equation}
Then using \rf{abcd}, \rf{QSeq} and \rf{ABeq}, we find
\begin{eqnarray}
\a=\frac{\zeta^{1/2}}{\sqrt{2g\g}}(A+B)\qquad&&\qquad\b=\frac{\zeta^{1/2}}{\sqrt{2g\g}}(A-B)\nonumber\\
\c=-\frac{\zeta^{-1/2}}{\sqrt{2g\g}}(A-B)\qquad&&\qquad\d=-\frac{\zeta^{-1/2}}{\sqrt{2g\g}}(A+B)\,.
\end{eqnarray}

However, we could also choose the  basis
\begin{equation}
 \QQ_{\al a}=2g\,\zeta^{1/2}U_{a\al}\qquad \SS_{a\al}=-2g\,\zeta^{-1/2}\left(\sqrt{\frac{\gamma^2}{16g^2}+1}+\frac{\gamma}{4g}(-1)^F\right)U_{a\al}\,.
 \end{equation}
 However, no matter the basis, we would always find
 \begin{equation}
 \SS_{a\al}=-\zeta^{-1}\left(\sqrt{\frac{\gamma^2}{16g^2}+1}+\frac{\gamma}{4g}(-1)^F\right)\QQ_{\al a}\,.
 \end{equation}
In the giant magnon regime, where $\frac{\gamma}{4g}<<1$, we see that 
\begin{equation}
 \SS_{a\al}\approx -\,\zeta^{-1}\QQ_{\al a} \,.
\end{equation}
 In the Metsaev-Tseytlin regime where $\frac{\gamma}{4g}>>1$, the relation between the generators is
 approximately
  \begin{equation}
 \ \SS_{a\al}\approx\zeta^{-1}\frac{\g}{4g}\left(1+(-1)^F\right)\QQ_{\al a}\,.
 \end{equation}

\renewcommand{\theequation}{3.\arabic{equation}}
 \setcounter{equation}{0}
 \section{Bosonic zero modes}
 
In this section we consider the bosonic zero modes for the giant magnon.  The zero modes are of two types.  The first corresponds to the broken $SO(4)$ symmetry that arises when choosing a direction for the $S^2$.  The $SO(4)$ isometry is broken to $SO(3)$, so we should expect 3 zero modes of this type.  There is also a zero mode due to the broken translation symmetry along the $x$ direction.  This zero mode can be found by mapping the corresponding zero mode in the sine-Gordon model back to the principle chiral model.

To find the zero modes corresponding to the broken $SO(4)$ symmetry, let us consider the target space coordinates for the $S^5$ in terms of the 6-dimensional vector $\vec n$, satisfying the constraint $\vec n\cdot\vec n=1$.  The bosonic action in conformal gauge is given by
\begin{equation}\label{bosaction}
\frac{\sqrt{\l}}{4\pi}\int d\xi d x\left(\dot{\vec n}\cdot\dot{\vec n}-\vec n\,'\cdot\vec n\,'-\Lambda(\vec n\cdot\vec n-1)\right)\,.
\end{equation}
The giant magnon solution has $n_4=n_5=n_6=0$, while the polar angles of the remaining $S^2$ are identified with $\theta$ and $\phi$.  Solving for the Lagrange multiplier $\Lambda$ gives
\begin{equation}
\Lambda=\dot{\vec n}\cdot\dot{\vec n}-\vec n\,'\cdot\vec n\,'=((\dot\phi)^2-(\phi')^2)\sin^2\th-(\th')^2\,,
\end{equation}
which for the giant magnon gives
\begin{equation}\label{Lameq}
\Lambda=\tanh^2x-\sech^2x\,.
\end{equation}

If we now consider the bosonic fluctuations about this solution, let us call $\vec Y$ the fluctuations along the $4$, $5$ and $6$ directions and $\vec N$ the fluctuations along the $1$, $2$ and $3$ directions.  Expanding to quadratic order in $\vec Y$ and $\vec N$, we find
\begin{equation}
\frac{\sqrt{\l}}{4\pi}\int d\xi d x\left(\dot{\vec Y}\cdot\dot{\vec Y}-\vec Y\,'\cdot\vec Y\,'-\Lambda\,\vec Y\cdot\vec Y+\dot{\vec N}\cdot\dot{\vec N}-\vec N\,'\cdot\vec N\,'-\Lambda\,\vec N\cdot\vec N\right)
\end{equation}
where $\Lambda$ is given in \rf{Lameq} and $\vec N$ satisfies 
\begin{equation}
\vec N\cdot \vec n_{GM}=0\,,
\end{equation}
 where $\vec n_{GM}$ is the giant magnon solution.  This last condition ends up coupling the different components of $\vec N$.
 
Zero-mode solutions for $\vec Y$ satisfy the equation
\begin{equation}
\vec Y''-\Lambda \vec Y=0\,,
\end{equation}
which has the three independent solutions
$\vec Y=\sech\,x\vec Y_0$ where $\vec Y_0$ is independent of $x$ and has components in directions $4$, $5$ and $6$.  These are the solutions that arise from the broken $SO(4)$ isometry and correspond to the freedom of choosing a transverse direction for the giant magnon.  The presence of these solutions also means that there are  canonical momenta that commute with the Hamiltonian (although not with each other), so there should be giant magnon solutions with different values for this momenta.    These correspond to the magnons with  a nonzero second $R$-charge, $Q$, discussed by Dorey.  The fermion zero modes are still present, so this means that these magnons will also have 8 bosonic and 8 fermionic states, and so there will be $16Q^2$ states in the full representation, as recently discussed in \cite{Chen:2006gp}.  

The zero mode coming from the broken translation symmetry is contained in $\vec N$.  One way to obtain this is to map the principle chiral model to the sine-Gordon theory, find the zero mode in sine-Gordon, and then map back to the principle chiral model.  The identification of the principle chiral model with the sine-Gordon theory has \cite{Pohlmeyer:1975nb,Mikhailov:2005qv,Mikhailov:2005sy}
\begin{equation}
\p_+\vec n\cdot\p_-\vec n=\cos2\Phi\,,
\end{equation}
where $\vec n=\vec n_{GM}+\vec N$.  The identification also requires the constraint equations
\begin{equation}
\p_+\vec n \cdot\p_+\vec n=1\,,\qquad\p_-\vec n \cdot\p_-\vec n=1\,.
\end{equation}
The soliton solution for sine-Gordon is
\begin{equation}\label{sgsol}
\Phi=\arccos\tanh x
\end{equation}
and the zero-mode solution is
\begin{equation}
\delta\Phi=\sech x\,,
\end{equation}
which is easily found by shifting $x$ in \rf{sgsol}.  Converting this back to the principle chiral model, one finds that this zero mode has the relatively simple form
\begin{equation}
N_3=-\tanh x\,\sech x\qquad\qquad N_1+iN_2=ie^{it}\,\sech^2 x\,.
\end{equation}

\renewcommand{\theequation}{4.\arabic{equation}}
 \setcounter{equation}{0}
 \section{Conclusions}
 
 In this paper we constructed the fermionic and bosonic zero modes for the Hofman-Maldacena giant magnon.  For the fermionic modes we explicitly constructed 8 zero modes and the solutions can be naturally fixed under kappa symmetry.  The final solutions are remarkably simple.  We also showed how to construct the generators for each $SU(2|2)$ superalgebra from the zero modes.
 
It would be of interest to find the fluctuation spectrum for the nonzero modes.  In this case the $\xi$ derivatives are no longer zero and this modifies the equation in \rf{compact}   to 
\begin{equation}\label{notascompact}
\frac{1}{\tanh x}\,(\DD-\p_\xi)\frac{1}{\tanh x}\,(\DD+\p_\xi)\Psi^1-\Psi^1=0\,.
\end{equation}
We have not yet succeeded in finding solutions for this equation for nonzero $\p_\xi\Psi^1$.

Another interesting direction concerns the higher order terms for the fluctuations.  The Green-Schwarz action normally has quartic terms in the world sheet action, even if the target space is flat ten dimensional space.  However, for flat space these higher order terms go away after fixing the kappa symmetry.  For the $AdS_5\times S^5$ background one can choose the gauge such there are at most quartic terms \cite{Metsaev:1998it,Metsaev:2000yf,Metsaev:2000yu}\footnote{I thank A. Tseytlin for clarifying remarks about these quartic terms.}. We believe that many of these terms will also be zero, at least for the zero modes.  However, it is not yet clear that all such terms will be removed.  

Ultimately, we would like to find the exact quantization of the Green-Schwarz string in the $AdS_5\times S^5$ background.  The classical integrability of the string theory \cite{Bena:2003wd}, the underlying superalgebra \cite{Beisert:2005tm,Beisert:2006qh} and the recent progress in \cite{Beisert:2006ez} hints that a full quantization can be obtained.  We believe that the results for the Hofman-Maldacena zero-modes  indicate that vast simplifications could occur that will allow one to make further advances in this direction.
 

\bigskip
\bigskip

\section*{Acknowledgments }

We thank A. Tseytlin and K. Zarembo for many helpful discussions.
 This work is supported in part by the Swedish Research Council and by the STINT foundation.  We also thank the CTP at MIT for hospitality during the course of this work.

\end{document}